\documentclass[a4paper,twoside]{article}

\usepackage{epsfig}
\usepackage{subcaption}
\usepackage{calc}
\usepackage{amssymb}
\usepackage{amstext}
\usepackage{amsmath}
\usepackage{amsthm}
\usepackage{multicol}
\usepackage{pslatex}
\usepackage{apalike}
\usepackage{url}

\usepackage{cite}
\usepackage{amsmath,amssymb,amsfonts}
\usepackage{algorithmic}
\usepackage{graphicx}
\usepackage{textcomp}
\usepackage{xcolor}
\usepackage{mathtools}
\usepackage{tikz}
\usetikzlibrary{shapes.multipart,shapes,shadows,arrows,decorations.markings,trees,positioning,decorations.markings,calc,fit,chains,intersections,decorations.pathreplacing}
\usepackage[underline=true]{pgf-umlsd}
\usepackage{flexisym}
\usepackage{siunitx}
\usepackage{float}
\usepackage{supertabular,booktabs}
\usepackage{enumitem,kantlipsum}
\usepackage{tcolorbox}
\usepackage{longtable}
\usepackage{tikz-qtree}
\usetikzlibrary{positioning, shapes.geometric}
\usepackage{arydshln}
\usepackage{upgreek}

\usepackage{SCITEPRESS}     

\begin{document}

\title{Privacy Enhanced DigiLocker using Ciphertext-Policy Attribute-Based Encryption}

\author{
	\authorname
	{
		Puneet Bakshi		
		and Sukumar Nandi	
	}
	\affiliation
	{
		Department of Computer Science and Engineering, Indian Institute of Technology, Guwahati, Assam, India
	}
	\email
	{
		\{b.puneet, sukumar\}@iitg.ac.in
	}
}


\abstract{Recently, Government of India has taken several initiatives to make India digitally strong such as to provide each resident a unique digital identity, referred to as \textit{Aadhaar}, and to provide several online e-Governance services based on Aadhaar such as \textit{DigiLocker}. DigiLocker is an online service which provides a shareable private storage space on public cloud to its subscribers. Although DigiLocker ensures traditional security such as data integrity and secure data access, privacy of e-documents is yet to addressed. Ciphertext-Policy Attribute-Based Encryption (CP-ABE) can improve data privacy but the right implementation of it has always been a challenge. This paper presents a scheme to implement privacy enhanced DigiLocker using CP-ABE.}

\onecolumn \maketitle \normalsize \setcounter{footnote}{0} \vfill

\section{\uppercase{Introduction}}\label{sec:introduction}
In last decade, Government of India has taken several e-Governance initiatives such as a unique digital identity (referred to as \textit{Aadhaar} \cite{aadhaar}) for every resident, online Aadhaar based authentication and several online citizen centric services such as \textit{eKYC}, \textit{eSign}, and \textit{DigiLocker}. At present, most of these services are built using traditional Public Key Infrastructure (PKI) with limited data privacy in which specifying authorized entities beforehand which are permitted to access data may not be possible and even if possible, the solution may not scale.

In \textit{DigiLocker} \cite{digilocker}, documents of subscribers are hosted on public cloud which is assumed to be a trusted entity. However, cloud storage may not be trustworthy and may be susceptible to insider attacks. Moreover, instead of providing a \textit{reactive access authorization to a single requester} (using OAuth2 \cite{oauth2}), a
 subscriber may want to provide a \textit{proactive access authorization to multiple requester} meeting certain criteria of attributes.

Ciphertext-Policy Attribute-Based Encryption (CP-ABE) \cite{bethencourt2007ciphertext} is a recent cryptographic mechanism which can improve data privacy, but the right implementation and efficiency of it are still some of the major concerns for its wide deployment. This paper presents a scheme to implement privacy enhanced DigiLocker based on CP-ABE.

\section{\uppercase{Related Work}}
\noindent Recent developments in cryptography have introduced Attribute-Based Encryption (ABE) \cite{goyal2006attribute} in which encryption is done under a set of attributes. ABE is classified in Key-Policy ABE (KP-ABE) \cite{goyal2006attribute} and CP-ABE. In KP-ABE, access policy is encoded in subscriber's private key and a set of attributes are encoded in ciphertext. In CP-ABE, access policy is encoded in ciphertext and a set of attributes are encoded in subscriber's private key. In CP-ABE, only if the set of required attributes encoded in receiver's private key satisfies the access policy encoded in received ciphertext, will the receiver be able to decrypt the ciphertext. Since the introduction of CP-ABE, researchers have proposed innovative mechanisms to use it to improve data privacy \cite{zhou2012efficient}, \cite{ji2014privacy}.

\section {\uppercase{Digital Lockers in India}}
\noindent \textit{DigiLocker} is an Aadhaar based online service which facilitates \textit{subscribers} to store e-documents, \textit{issuer} agencies to provide e-documents and \textit{requester} applications to get access to e-documents. An \textit{e-document} is a digitally signed electronic document. \textit{Repositories} are provided by issuers to host collection of e-documents. \textit{Digital Locker} is a storage space provided to each subscriber to store e-documents. \textit{Requester} is an application which seeks access to some e-document. All participating entities must adhere to \textit{Digital Locker Technology Specification (DLTS)} \cite{dlts}. 

An e-document is uniquely identified by a \textit{Unique Resource Identifier (URI)} which is a triplet of the form $\mathrm{\langle{IssuerID::DocType::DocID}\rangle}$, where $IssuerID$ is a unique identifier ot the issuer, e.g., $\mathtt{CBSE}$, for Central Board of Secondary Education. $DocType$ is a classification of e-documents as defined by the issuer. For example, CBSE may classify certificates into $\mathtt{MSTN}$ for 10th mark sheet and $\mathtt{KVYP}$ for certificates issued to KVPY scholarship fellow. $DocType$ also helps issuers to use different repositories for different types of e-documents. $DocID$ is an issuer defined unique identifier (an alphanumeric string) of the e-document within a document type. Some hypothetical examples of e-document URI are $\langle{\mathtt{CBSE::MSTN::22636726}}\rangle$, $\langle{\mathtt{DLSSB::HSMS::GJSGEJXS}\rangle}$. DigiLocker ensures data integrity of e-documents by mandating that all e-documents are digitally signed by issuers.

When an issuer is registered, it provides two APIs, namely, $\mathtt{PullDoc}$ to pull an e-document based on a given URI and $\mathtt{PullUri}$ to pull all URIs meeting a given search criteria. When a requester application is registered, it is given a unique requester identifier, a secret key which is shared between DigiLocker and requester application and a $\mathtt{FetchDoc}$ API is given to access e-documents based on URI. Based on the URI, $\mathtt{FetchDoc}$ forwards the request to appropriate issuers to retrieve the e-document. DigiLocker ensures secure data access of e-documents by API license keys, secure transport, an explicit authentication (if required by \textit{DocType}) and all requests and responses to be digitally signed.

\section{\uppercase{Preliminaries}}
\noindent This section briefly describes some of the necessary background.

\subsection{Bilinear pairings \cite{zhang2004efficient}}
Let $\mathbb{G}_1$ and $\mathbb{G}_2$ are elliptic groups of order $\mathrm{p}$, $\mathbb{G}_{\mathrm{T}}$ is a multiplicative group of order $\mathrm{p}$, $\mathrm{g_1}$ is a generator of $\mathbb{G}_{1}$, $\mathrm{g_2}$ is a generator of $\mathbb{G}_2$, $\mathrm{P} \in \mathbb{G}_1$, $\mathrm{Q}\in \mathbb{G}_2$ and $\mathrm{a}$, $\mathrm{b}$ $\in \mathbb{Z}_p$, then a bilinear pairing is a map $\mathrm{e}: \mathbb{G}_1 \times \mathbb{G}_2 \rightarrow \mathbb{G}_{\mathrm{T}}$ that satisfies the following three properties.
\begin{enumerate}
	\item [1] Bilinearity: $\mathrm{e(P^a, Q^b) = e(P, Q)^{ab}}$
	\item [2] Non-Degeneracy: $\mathrm{e(g_1, g_2) \neq 1}$
	\item [3] Computability: $\mathrm{e(P, Q)}$ can be computed efficiently.
\end{enumerate}

\subsection {Decision Bilinear Diffie-Hellman (DBDH) assumption \cite{yacobi2002note}}
Let $\mathbb{G}$, $\mathbb{G}_T$ are cyclic groups of prime order $\mathrm{p} > 2^{\lambda}$ where $\lambda \in \mathbb{N}$, $\mathrm{g}$ is the generator of $\mathbb{G}$, $\mathrm{e}: \mathbb{G} \times \mathbb{G} \rightarrow \mathbb{G}_{\mathrm{T}}$ is an efficiently computable symmetric bilinear pairing map and $\mathrm{a, b, c, z} \in \mathbb{Z}_{\mathrm{p}}$ are random numbers. The DBDH assumption states that no probabilistic polynomial time algorithm can distinguish between $\mathrm{\langle{{g},{g^a},{g^b},{g^c},{e(g,g)^{abc}}}\rangle}$ and $\mathrm{\langle{{g},{g^a},{g^b},{g^c},{e(g,g)^{z}}}\rangle}$ with more than a negligible advantage.

\subsection {Security Model}
The security model of proposed scheme is based on the following IND-sAtt-CPA game \cite{ibraimi2009efficient} between a challenger and an adversary $\mathcal{A}$. 
\begin{enumerate}[leftmargin=0pt]
	\item []\textit{Init Phase}: Adversary $\mathcal{A}$ chooses a challenge access tree $\mathcal{T}^*$ and gives it to challenger.
	
	\item []\textit{Setup Phase}: Challenger runs a \textit{setup} procedure to generate $\mathrm{\langle{ASK, APK}\rangle}$ and gives the public key $\mathrm{APK}$ to adversary $\mathcal{A}$.
	
	\item []\textit{Phase I}: Adversary $\mathcal{A}$ makes an attribute- based private key request to the key generation oracle for any attribute set with the restriction that the attribute set should not include any attribute which is part of $\mathcal{T}^*$. Challenger generates the key as described in section \ref{key_generation} and returns the same to adversary $\mathcal{A}$.
	
	\item []\textit{Challenge Phase}: Adversary $\mathcal{A}$ sends two equal length messages $\mathrm{m_0}$ and $\mathrm{m_1}$ to challenger. Challenger chooses a random number $\mathrm{b} \in_{\mathrm{R}} \{0, 1\}$, encrypts $\mathrm{m_b}$ using $\mathcal{T}^*$ and $\mathrm{APK}$ as is described in section \ref{encryption}.
	
	\item []\textit{Phase II}: Adversary $\mathcal{A}$ can send multiple requests to generate attribute-based private key with the same restriction as in Phase I.
	
	\item []\textit{Guess Phase}: Adversary $\mathcal{A}$ outputs a guess $\mathrm{b\prime} \in \{0, 1\}$.
\end{enumerate}
The advantage of adversary $\mathcal{A}$ in this game is defined to be $\epsilon = \vert {\mathrm{Pr[b\prime = b]} - \frac{1}{2}} \vert$. Only if any polynomial time adversary $\mathcal{A}$ has a negligible advantage, the scheme is considered secure against an adaptive chosen plaintext attack (CPA).

\subsection {Access tree} \label{access_tree}
Access tree structure is a means to specify an access policy during encryption that must be satisfied by attribute-based private keys in order to decrypt. Let $\mathcal{T}$ be a tree representing an access structure. Each non-leaf node of the tree represents a threshold gate, described by its children and a threshold value. If $\mathrm{num_x}$ is the number of children and $\mathrm{k_x}$ is the threshold value of a node $\mathrm{x}$, then, $\mathrm{k_x=1}$ represents an OR gate and $\mathrm{k_x=num_x}$ represents an AND gate. Each leaf node $\mathrm{x}$ of the tree is described by an attribute and a threshold value $\mathrm{k_x = 1}$. 

Let $\mathcal{T}_x$ denotes the subtree rooted at node $\mathrm{x}$. If a set of attributes $\lambda$ satisfies the subtree $\mathcal{T}_x$, it is represented as $\mathcal{T}_x(\lambda) = 1$. $\mathcal{T}_x(\lambda)$ is computed recursively as follows. If $\mathrm{x}$ is a non-leaf node, evaulate $\mathcal{T}(y)$ for all children nodes $\mathrm{y}$ of node $\mathrm{x}$. $\mathcal{T}_x(\lambda)$ returns $1$ if and only if at least $\mathrm{k_x}$ children return 1. If $\mathrm{x}$ is a leaf node, then $\mathcal{T}_x(\lambda)$ returns $1$ if and only if $\mathtt{attr(x)} \in \lambda$.

\section {\uppercase{Our Construction}}

\noindent The proposed scheme introduces two new roles, namely, \textit{Attribute Authority Manager (AAM)} and \textit{Attribute Authority (AA)}. AAM is an entity which manages the universe of attributes and AA is an entity which manages a set of attributes (as assigned by AAM). DigiLocker is proposed to assume the role of AAM and individual issuers are proposed to assume the role of AA. A subscriber is assigned a set of attributes from each issuer which holds at least one e-document of the subscriber. Each requester application is assigned a set of attributes from DigiLocker based on certain criteria such as purpose of access, for how long the data is going to be used, etc. To create a privacy enhanced e-document for a subscriber, issuer and subscriber mutually creates an attribute-based token (which will be used later in encryption) for an access policy, generates a symmetric key, encrypts the document with symmetric key, encrypts the symmetric key with attribute-based token, creates an e-document enclosing both the encrypted symmetric key and the encrypted document, creates a URI for this e-document and pushes it to subscriber's digital locker using $\mathtt{PushURI}$ API. When this e-document is shared with a requester application, the requester will be able to decrypt the encrypted symmetric key only if the requester is associated with a set of attributes which satisfies the access policy used to encrypt the symmetric key. Only when the requester obtains the symmetric key, will he be able to decrypt and retrieve the document.

In \textit{Setup($\kappa$)} procedure, AAM chose a cyclic group ${\mathbb{G}_0}$ of large prime order $\mathrm{p}$ ($\kappa$ defines the size of group) on which discrete logarithm problem is assumed to be hard, generator $\mathrm{g}$, a bilinear map $\mathrm{e}:{\mathbb{G}_0}\times {\mathbb{G}_0}\rightarrow {\mathbb{G}_1}$ for which bilinear diffie hellman problem is assumed to be hard, a hash function $\mathrm{H}: \{0, 1\}^{*} \rightarrow \mathbb{G}_0$ which maps a binary string encoded attribute to a group element, chose random numbers $\alpha, \beta \in_{\mathrm{R}} \mathbb{Z}_{\mathrm{p}}$ and set its private key $\mathrm{ASK}$ and public key $\mathrm{APK}$ as below.
\begin{center}
	\begin{tabular}{ l c l }
		$\mathrm{ASK}$ & = & $\{\beta, \mathrm{g}^{\alpha} \}$ \\ 
		$\mathrm{APK}$ & = & $\{\mathrm{g}^{\beta}, \mathrm{e(g, g)}^{\alpha}, \mathbb{G}_0, \mathrm{g}\}$ \\  
	\end{tabular}
\end{center}
\subsection {Attribute Assignment}{\label{attribute_assignment}}
An attribute can be any characteristic of a subscriber or requester and is represented by a binary string $\{0,1\}^*$. Attribute assignment to both subscribers and requesters is proposed to be done lazily in the background with the aim to keep the list of associated attributes in DigiLocker up to date.

For subscriber's attribute assignment and modification, two APIs are proposed to be introduced. First is  $\mathtt{PullAttrs(ID_i)}$ which is provided by issuers and is consumed by DigiLocker to pull updated list of attributes of subscriber with Aadhaar number $\mathrm{ID_i}$. Second is $\mathtt{PushAttrs(ID_i, NewAttrs)}$ which is provided by DigiLocker and is consumed by issuer to push any change in attributes of subscriber with Aadhaar number $\mathrm{ID_i}$. For requester applications, attributes are assigned and updated by DigiLocker.

It is important to take appropriate measures to handle load of a voluminous country like India. One such measure could be to prepone part of the encryption process. This preponed encryption process generates a token with mutual cooperation between subscriber and issuer. This token can be reused every time for a given subscriber and for a given access policy. 

A helper procedure $\mathtt{encPartial(\mathcal{T}, r)}$ is assumed to be present which works as follows. It chooses a polynomial $\mathrm{q_x}$ for each node $\mathrm{x}$ (including the leaves) in the tree $\mathcal{T}$. These polynomials are chosen in the following way in a top-down manner, starting from the root node $\mathrm{R}$. For each node $\mathrm{x}$ in the tree, set the degree $\mathrm{d_x}$ of the polynomial $\mathrm{q_x}$ to be one less than the threshold value $\mathrm{k_x}$ of that node, that is, $\mathrm{d_x = k_x - 1}$. Starting with the root node $\mathrm{R}$ the procedure chooses a random $\mathrm{r}\in_{\mathrm{R}} \mathbb{Z}_{\mathrm{p}}$ and sets $\mathrm{q_r(0)=r}$. Then, it chooses $\mathrm{d_R}$ other points of the polynomial $\mathrm{q_R}$ randomly to define it completely. For any other node $\mathrm{x}$, it sets $\mathrm{q_x(0)} = \mathrm{q_{parent(x)}(index(x))}$ and chooses $\mathrm{d_x}$ other points randomly to completely define $\mathrm{q_x}$.

\subsection {Token Generation}{\label{token_generation}}
\begin{center}
\begin{figure}
	\caption{Example of an access policy tree}
	\label{fig:AP}
	\resizebox {0.9\columnwidth}{!} {
		\begin{tikzpicture} [level	distance = 50pt,
		sibling distance = 20pt, 
		edge from parent/.style = {
			draw, 
			edge from parent path = {(\tikzparentnode) -- (\tikzchildnode.north)}
		}
		]
		\tikzset{every internal node/.style	= {draw, circle, black, font = \Large}}
		\tikzset{every leaf node/.style		= {draw, black, regular polygon, regular polygon sides = 3, inner sep = 1pt}}
		
		\Tree [.\node(1)[label={[label distance=0.25cm]10:\LARGE $R$}]{$\lor$};
		[.\node(2)[label={[label distance=0.25cm]10:\LARGE ${R_S}_{iv}$}]{$\wedge$};
		\node(3){${A_R}_1$};
		\node(4){${A_R}_n$};
		] 
		[.\node(5)[label={[label distance=0.25cm]10:\LARGE ${R_I}_{iv}$}]{$\wedge$};
		\node(6){${A_R}_1$};
		\node(7){${A_R}_n$};
		] 
		]
		
		\node[draw, dashed, inner xsep=3mm, inner ysep=6mm, fit=(2)(3)(4)](r1){};
		\node[draw, dashed, inner xsep=3mm, inner ysep=6mm, fit=(5)(6)(7)](r2){};
		\node[draw, dashed, inner xsep=7mm, inner ysep=9mm, fit=(3)(1)(7)](r3){};
		
		\node [above, inner sep=5pt] at (r1.south) {\large ${\mathcal{T}_S}_{iv}$};
		\node [above, inner sep=5pt] at (r2.south) {\large ${\mathcal{T}_I}_{iv}$};
		\node [above, inner sep=5pt] at (r3.south) {\large ${\mathcal{T}_{iv}}$};
		
		
		\end{tikzpicture}
	}
\end{figure}
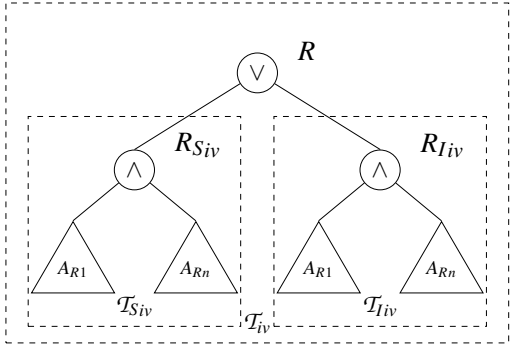
\end{center}

An access tree $\mathcal{T}_{iv}$ is comprised of access subtree $\mathcal{T}_{S_{iv}}$ from subscriber $\mathrm{S_i}$ and access subtree $\mathcal{T}_{I_{iv}}$ from issuer $\mathrm{I_v}$ (refer figure \ref{fig:AP}). If issuer $\mathrm{I_v}$ needs to generate its part of token for subscriber $\mathrm{S_i}$, for access tree $\mathcal{T}_{iv}$, it generates a random number $\mathrm{r_i} \in_{\mathrm{R}} \mathbb{Z}_{\mathrm{p}}$, and generates following partial-token using $\mathrm{APK}$ and $\mathtt{encPartial(\mathcal{T}_{I_{iv}}, r_i)}$. Let $\mathrm{Y_I}$ is the set of leaf nodes in $\mathcal{T}_{I_{iv}}$.

$\mathrm{CTtok_{I_{iv}}} =	\begin{cases}
	\quad \mathcal{T}_{I_{iv}}\\
	\quad \mathrm{C1_I} = \mathrm{e(g, g)}^{\alpha \mathrm{r_i}}\\
	\quad \mathrm{C2_I} = \mathrm{g}^{\beta \mathrm{r_i}}\\
	\begin{rcases}
		\quad \mathrm{C3_{I_y} = g^{q_y(0)}}\\
		\quad \mathrm{C4_{I_y} = H(\mathtt{attr(y)})^{q_y(0)}}\\
	\end{rcases} {\!\scriptscriptstyle\mathrm{\forall\ y \in Y_I}}\\
\end{cases}$
Issuer notifies subscriber to provide its part of the token. Subscriber $\mathrm{S_i}$ generates a random number $\mathrm{r_s} \in_{\mathrm{R}} \mathbb{Z}_{\mathrm{p}}$ and generates following partial-token using $\mathrm{APK}$ and $\mathtt{encPartial(\mathcal{T}_{S_{iv}})}$. Let $\mathrm{Y_S}$ is the set of leaf nodes in $\mathcal{T}_{S_{iv}}$.

$\mathrm{CTtok_{S_{iv}}} = \begin{cases}
	\quad \mathcal{T}_{S_{iv}}\\
	\quad \mathrm{C1_S} = \mathrm{e(g, g)}^{\alpha \mathrm{r_s}}\\
	\quad \mathrm{C2_S} = \mathrm{g}^{\beta \mathrm{r_s}}\\
	\begin{rcases}
		\quad \mathrm{C3_{S_y} = g^{q_y(0)}}\\
		\quad \mathrm{C4_{S_y} = H(\mathtt{attr(y)})^{q_y(0)}}\\
	\end{rcases} {\!\scriptscriptstyle\mathrm{\forall\ y \in Y_S}}\\
\end{cases}$

Subscriber provides its part of partial-token to issuer. Issuer creates the final token by combining the two partial-tokens and keeps it securely with it.

$\mathrm{CTtok_{iv}} = \begin{cases}
	\quad \mathcal{T}_{iv} = \mathcal{T}_{S_{iv}} \cup \mathcal{T}_{I_{iv}}\\
	\quad \mathrm{C1} = \mathrm{C1_S}.\mathrm{C1_I}\\
	\qquad\ \ = \mathrm{e(g, g)}^{\alpha \mathrm{r_s}}\mathrm{e(g, g)}^{\alpha \mathrm{r_i}}\\
	\quad \mathrm{C2} = \mathrm{C2_S}.\mathrm{C2_I} = \mathrm{g}^{\beta \mathrm{r_s}}\mathrm{g}^{\beta \mathrm{r_i}}\\
	\begin{rcases}
		\quad \mathrm{C3 = C3_{S_y} \cup C3_{I_y}}\\
		\qquad\ \  = \mathrm{g^{q_y(0)}}\\
		\quad \mathrm{C4 = C4_{S_y} \cup C4_{I_y}}\\
		\qquad\ \ = \mathrm{H(\mathtt{attr(y)})^{q_y(0)}}\\
	\end{rcases} {\!\scriptscriptstyle\forall\ y \in Y_S \cup Y_I}\\
\end{cases}$

\subsection{Encryption} {\label {encryption}}
A new $DocType$ $\mathtt{PRIV}$ is proposed to be introduced for privacy enhanced e-documents. To create a privacy enhanced e-document, issuer creates a URI $\mathrm{\langle{I_v::PRIV::D_w}\rangle}$ where $\mathrm{I_v}$ is the issuer identifier and $\mathrm{D_w}$ is the document identifier within the document type $\mathtt{PRIV}$. Now, issuer generates a random number $\mathrm{r_{ie}} \in_{\mathrm{R}} \mathbb{Z}_{\mathrm{p}}$, generates a symmetric key $\mathrm{SK_{ivw}}$, encrypts e-document $\mathrm{m}$ with $\mathrm{SK_{ivw}}$, encrypts $\mathrm{SK_{ivw}}$ with $\mathrm{CTtok_{I_{iv}}}$ and produces the following ciphertext.

$\mathrm{CT_{ivw}} = \begin{cases}
	\quad \mathcal{T}_{iv} = \mathcal{T}_{S_{iv}} \cup \mathcal{T}_{I_{iv}}\\
	\quad \mathrm{C1} = \mathrm{e(g, g)}^{\alpha \mathrm{r_s r_{ie}}}\mathrm{e(g, g)}^{\alpha \mathrm{r_i r_{ie}}}\mathrm{SK_{ivw}}\\
	\quad \mathrm{C2} = \mathrm{g}^{\beta \mathrm{r_s r_{ie}}}\mathrm{g}^{\beta \mathrm{r_i r_{ie}}}\\
	\begin{rcases}
		\quad \mathrm{C3_{y} = g^{r_{ie}(q_y(0))}}\\
		\quad \mathrm{C4_{y} = H(\mathtt{attr(y)})^{q_y(0)}}\\
	\end{rcases} {\!\scriptscriptstyle\mathrm{\forall\ y \in Y_{iv}}}\\
	\quad \mathrm{C5 = \{m\}_{SK_{ivw}}}\\
\end{cases}$

\subsection {Key Generation} {\label{key_generation}
A new API $\mathtt{GenABPvtKey(ID_i, IS_j)}$ is proposed to be provided by DigiLocker to generate an attribute-based private key for a subscriber with Aadhaar identifier $\mathrm{ID_i}$ and with attributes from issuers in the set $\mathrm{IS_j}$. Let $\mathrm{S_{ij}}$ is the set of all attributes assigned to $\mathrm{S_i}$ by all issuers in set $\mathrm{IS_j}$. DigiLocker generates random numbers $\mathrm{r} \in_{\mathrm{R}} \mathbb{Z}_{\mathrm{p}}$, $\mathrm{r_j} \in_{\mathrm{R}} \mathbb{Z}_{\mathrm{p}}$ for each attribute $\mathrm{j}\in_{\mathrm{R}} \mathrm{S_{ij}}$, computes attribute-based private key $\mathrm{ASK_{ID_{i}IS_{j}}}$ as below and keeps this key securely with it.

	$\mathrm{ASK_{{ID_i}{IS_j}}} = \begin{cases}
		\quad \mathrm{D} = \mathrm{g}^{(\alpha+\mathrm{r})/\beta}\\
		\begin{rcases}
			\quad \mathrm{D_j} = \mathrm{g^{r}.H(j)^{r_j}}\\
			\quad \mathrm{D_j\prime} = \mathrm{g^{r_j}}\\
		\end{rcases} {\!\scriptscriptstyle\mathrm{\forall\ j \in S_{ij}}}\\
	\end{cases}$

Note that multiple attribute based private keys can exist for a subscriber for different set of attributes. If anyone issuer set $\mathrm{IS_i}$ is a proper subset of another issuer set $\mathrm{IS_j}$, the key corresponding to $\mathrm{IS_i}$ is redundant and can be removed.

\subsection {Decryption} {\label{decryption}}
A new API $\mathtt{FetchPrivDoc{URI}}$ is proposed to be provided by DigiLocker for decryption purpose. This API facilitates a requester with identifier $\mathrm{ID_R}$ to retrieve ciphertext $\mathrm{CT_{ivw}}$ of e-document from URI $\mathrm{\langle{I_v::PRIV::D_W}\rangle}$ of subscriber $\mathrm{S_i}$. DigiLocker extracts the set of attribute issuers $\mathrm{IS_k}$ from $\mathrm{CT_{ivw}}\rightarrow\mathcal{T}_{iv}$, retrieves $\mathrm{ASK_{ID_RIS_k}}$ and calls $\mathtt{Decrypt(CT_{ivw}, ASK_{ID_RIS_k})}$. A helper procedure $\mathtt{DecryptNode(CT_{ivw}, ASK_{ID_RIS_K})}$ is defined as below. Let $\mathrm{S_k}$ is the set of all attributes from issuers in set $\mathrm{IS_k}$. If $\mathrm{x}$ is a leaf node and if $\mathtt{attr(x)} \notin \mathrm{S_k}$, then $\mathtt{DecryptNode(CT_{ivw}, ASK_{{ID_R}{IS_k}}, x)} = \perp$ else if $\mathtt{attr(x)}\in \mathrm{S_k}$, then the procedure is defined as below.
\begin{tabular}{ m{1cm} l }
	\multicolumn{2}{l}{ $\mathtt{DecryptNode(CT_{ivw}, ASK_{{ID_R}{IS_k}}, x)}$ }\\
	& = $ \mathrm{\frac{e(D_x, C4_y)}{e(D_x\prime, C5_y)}}$ \\ 
	& = $ \mathrm{\frac{e(g^r.H(\mathtt{attr(x)})^{r_j}, g^{r_{ie}q_y(0)})}{e(g^{r_j}, H(\mathtt{attr(x)})^{q_x(0)})}}$ \\  
	& = $ \mathrm{e(g, g)^{rr_{ie}q_x(0)}}$
\end{tabular}

If $\mathrm{x}$ is a non-leaf node, the recursive procedure is defined as follows. For all children nodes $\mathrm{z}$ of $\mathrm{x}$, $\mathtt{DecryptNode(CT_{ivw}, ASK_{i}, x)}$ is called and their output is stored in $\mathrm{F_z}$. Let $\mathrm{S_x}$ be an arbitrary $\mathrm{k_x}$ sized set of child nodes $\mathrm{z}$ such that $\mathrm{F_z} \neq \perp$. If no such set exists then the node was not satisfied and the function returns $\perp$. Otherwise, $\mathrm{F_x}$ is computed as below.

\begin{tabular}{ m{0cm} l }
	\multicolumn{2}{l}{$\mathrm{F_x}$ = $  \mathrm{\displaystyle\prod_{z\in S_x} F_{z}^{\Delta_{i, S_x\prime}(0)}\quad {\tiny where \{ {i=index(z)} \atop  {S_x\prime = \{index(z): z \in S_x} \} }} $ }\\
	& = $ \mathrm{\displaystyle\prod_{z\in S_x} F_{z}^{\Delta_{i, S_x\prime}(0)}}$ \\ 
	& = $ \mathrm{\displaystyle\prod_{z\in S_x} (e(g, g)^{r.r_{ie}.q_z(0)})^{\Delta_{i, S_x\prime}(0)}}$\\  
	& = $ \mathrm{\displaystyle\prod_{z\in S_x} (e(g, g)^{r.r_{ie}.q_{parent(z)}(index(z))})^{\Delta_{i, S_x\prime}(0)}}$\\
	& = $ \mathrm{\displaystyle\prod_{z\in S_x} e(g, g)^{r.r_{ie}.q_x(i)\Delta_{i, S_x\prime}(0)}}$\\
	& = $ \mathrm{e(g, g)^{rr_{ie}q_x(0)}\ (\tiny{using\ polynomial\ interpolation})}$
\end{tabular}
$\mathtt{Decrypt}(\mathtt{CT_{ivw}}, \mathtt{ASK_{{ID_R}{IS_k}}})$ calls $\mathtt{DecryptNode(}$ $\mathtt{CT_{ivw}}, \mathtt{ASK_{{ID_R}{IS_k}}}, \mathtt{R})$ where $\mathrm{R}$ is root node of $\mathrm{{T}_{iv}}$. If the access tree is satisfied by attributes in $\mathrm{S_k}$, set $\mathrm{A =}$ $\mathtt{DecryptNode(CT_{ivw}, ASK_{{ID_R}{IS_k}}, R)} =$ $\mathrm{e(g, g)}^{\mathrm{r}.\mathrm{r_{ie}}.(\mathrm{r_s}+\mathrm{r_i})}$. Now the procedure obtains symmetric key $\mathrm{SK_{ivw}}$ by computing 
\begin{tabular}{ m{1cm} l }
	\multicolumn{2}{l}{$\mathrm{\dfrac{C1}{\dfrac{e(C2, D)}{A}}}$ = $\dfrac{\mathrm{e(g,g)}^{\alpha \mathrm{r_s r_{ie}}}\mathrm{e(g,g)}^{\alpha \mathrm{r_i r_{ie}}}\mathrm{SK_{ivw}}}{\dfrac{\mathrm{e}(\mathrm{g}^{\beta \mathrm{r_s r_{ie}}}\mathrm{g}^{\beta \mathrm{r_i r_{ie}}}, \mathrm{g}^{(\alpha+\mathrm{r})/\beta})}{\mathrm{e(g, g)}^{\mathrm{r r_{ie} (r_s+r_i) )} }}}$}\\
	& = $\dfrac{\mathrm{e(g,g)}^{\alpha \mathrm{r_s r_{ie}}}\mathrm{e(g,g)}^{\alpha \mathrm{r_i r_{ie}}}\mathrm{SK_{ivw}}} {\dfrac {\mathrm{e}(\mathrm{g}^{\beta \mathrm{r_{ie} (r_s+r_i)}}, \mathrm{g}^{(\alpha+\mathrm{r})/\beta})} {\mathrm{e(g, g)}^{\mathrm{r r_{ie} (r_s+r_i))}}}}$ \\ 
	& = $\dfrac{\mathrm{e(g,g)}^{\alpha \mathrm{r_{ie} (r_s + r_i)}}\mathrm{SK_{ivw}}} {\dfrac {\mathrm{e(g, g)}^{\mathrm{r_{ie} (r_s+r_i)} (\alpha+\mathrm{r})}} {\mathrm{e(g, g)}^{\mathrm{r r_{ie} (r_s+r_i))}}}}$\\  
	& = $\mathrm{SK_{ivw}}$\\
\end{tabular}

Symmetric key $\mathrm{SK_{ivw}}$ is now used to decrypt the encrypted e-document.

\begin{tabular}{m{0cm} l}
	\multicolumn{2}{l}{$\mathrm{m = \{CT_{ivw}\rightarrow{C5}\}_{SK_{ivw}}}$}
\end{tabular}

DigiLocker returns the decrypted document $\mathrm{m}$ to requester.

\section{\uppercase{Security Analysis}}
\noindent If the proposed scheme is not secure than an adversary $\mathcal{A}$ can win IND-sAtt-CPA game and solve the DBDH assumption with advantage $\epsilon/2$. If the DBDG assumption is solved by adversary $\mathcal{A}$, a simulator $\beta$ can be built which can solve DBDH assumption with advantage $\epsilon/2$. Challenger chose a group $\mathbb{G}_0$, a generator $\mathrm{g}$, a bilinear map $\mathrm{e}$ and chose random numbers $\mathrm{a, b, c,} \theta \in_{\mathrm{R}} \mathbb{Z}_p^*$. The challenger selects at random $\mu \in_{\mathrm{R}} {0, 1}$ and sets $\mathrm{Z}_\mu$ as below.
\begin{align*}
	\mathrm{Z}_\mu = \begin{cases}
	\mathrm{(g, g)^{abc}}, &\mathrm{if}\ \mu = 0\\
	\mathrm{e(g, g)}^{\theta}, &\mathrm{if}\ \mu = 1\\
	\end{cases}	
\end{align*}

Challenger provides DBDB challenge to the simulator: $\langle{\mathrm{g, A, B, C, Z}_{\mu}}\rangle$  $\langle{\mathrm{g, g^a, g^b, g^c, Z}_{\mu}}\rangle$. In IND-sAtt-CPA game, simulator $\beta$ plays the role of challenger for adversary $\mathcal{A}$.
	
\begin{itemize}[leftmargin=*]
		\item [] \textit{Init Phase}: The adversary chose the challenge access tree $\mathcal{T}^{*}$ and gives it to simulator.
		
		\item [] \textit{Setup Phase}: The challenger chose a random number $\mathrm{x\prime} \in \mathbb{Z}_p$, sets $\alpha = \mathrm{ab + x\prime}$ and computes $\mathrm{y}$ as below.
		\[
		\mathrm{y = e(g, g)}^{\alpha} = \mathrm{e(g, g)^{ab}e(g, g)^{x\prime}}
		\]
		Now, challenger chose a random numbers $\mathrm{r \in_R \mathbb{Z}_p}$ and $\mathrm{r_i \in_R \mathbb{Z}_p}$ for ($\mathrm{1 \le i \le \vert{\mathtt{U}}\vert}$) and for all $\mathrm{a_j \in \mathbb{U}}$, computes $\mathrm{d_j}$ and $\mathrm{d_j\prime}$ as below.
		\begin{align*}
			\begin{rcases}
				\mathrm{d_j} &= \begin{cases}
					\mathrm{g^{r/b} H(j)^{r_j}} & \mathrm{...if\ a_j} \notin \mathcal{T}^*\\
					\mathrm{g^{r} H(j)^{r_j}} & \mathrm{...if\ a_j} \in \mathcal{T}^*\\
					\end{cases}\\
					\mathrm{d_j\prime} &= \mathrm{g^{r_j}}
				\end{rcases} 
			\mathrm{(1 \le j \le \vert{U}\vert)}
		\end{align*}
		Now, challenger sends public parameters $\mathrm{APK} = \{\mathrm{g}^\beta, \mathrm{e(g, g)}^\alpha, \mathbb{G}, \mathrm{g}\}$ to adversary $\mathcal{A}$.
		
		\item [] \textit{Phase 1}: In this phase, adversary $\mathcal{A}$ sends requests for private key for any set of attributes $w_j$ which does not contain any attribute in $\mathcal{T}^*$.
		\begin{align*}
			\mathrm{w_j} = \{\mathrm{a_j} \mid (\mathrm{a_j} \in \mathbb{U} \wedge \mathrm{a_j} \notin \mathcal{T}^*)\}
		\end{align*}
		For each query from adversary $\mathcal{A}$, challenger chose a random number $\mathrm{r\prime \in_R\mathbb{Z}_p}$, sets $\mathrm{r= - b(r\prime + a)}$ and computes $\mathrm{D}$ as below.
		\begin{align*}
			\mathrm{D} &= \mathrm{g}^{(\alpha+\mathrm{r})/\beta} = (\mathrm{g}^{(\alpha+\mathrm{r})})^{1/\beta} = (\mathrm{g}^{((\mathrm{ab+x\prime})-\mathrm{b(r\prime+a)})})^{1/\beta}\\
			&= (\mathrm{g^{x\prime-br\prime})}^{1/\beta} = (\mathrm{g^{x\prime}.(g^{b})^{-r\prime})}^{1/\beta}
		\end{align*}
		Because of restriction $\mathrm{a_j} \notin \mathcal{T}^*$ in this phase, $\mathrm{D_j}$ can be computed as below.
		\begin{align*}
			\mathrm{D_j} &= \mathrm{g^{r/b}H(j)^{r_j} = g^{r/b}H(j)^{r_j} = g^{-(r_\prime+a)}H(j)^{r_j}}\\
			    &= \mathrm{(g^{a})^{-1}g^{-r\prime}H(j)^{r_j}}
		\end{align*}
		Now, challenger sends private key $\mathrm{ASK_{w_j}} = \mathrm{D}, \mathrm{(D_j, D_j\prime)}\mid\forall \mathrm{a_j\in w_j}$ to adversary $\mathcal{A}$
		
		\item [] \textit{Challenge Phase}: In this phase, adversary $\mathcal{A}$ submits two plaintext messages $\mathrm{m_0}$ and $\mathrm{m_1}$ to the challenger. Challenger selects a random plaintext message $\mathrm{m_b}$ from the two messages where $\mathrm{b \in_R \{0, 1\}}$, sets $\mathrm{r_{ie}=1}$, chose random variables $\mathrm{r_i}$ and $\mathrm{r_s}$ such that $\mathrm{c=r_i+r_c}$. Now, set value of root node $\mathcal{T}^*$ to $\mathrm{c}$ and assign values to leaf nodes of $\mathcal{T}^*$ as described in section \ref{access_tree} to arrive at $\mathrm{C3_y}$ and $\mathrm{C4_y}$. The final ciphertext $\mathrm{CT}_{\mathcal{T}^*}$ is computed as below. The ciphertext is returned to adversary $\mathcal{A}$.
		
		$CT_{\mathcal{T}^*} = \begin{cases}
			\quad \mathcal{T}_{iv} = \mathcal{T}^*\\
			\quad \mathrm{C1} = \mathrm{e(g, g)}^{\alpha \mathrm{r_s}}\mathrm{e(g, g)}^{\alpha \mathrm{r_i}}\mathrm{m_b}\\
			\qquad\ \  = \mathrm{e(g, g)}^{\alpha(\mathrm{r_s+r_i})}\mathrm{m_b}\\
			\qquad\ \  = \mathrm{e(g, g)^{c}m_b}\\
			\quad \mathrm{C2} = \mathrm{g}^{\beta \mathrm{r_s}}\mathrm{g}^{\beta \mathrm{r_i}} = \mathrm{g}^{\beta(\mathrm{r_s+r_i})}\\
			\qquad\ \  = \mathrm{g}^{\beta} \mathrm{g^{c}}\\
			\begin{rcases}
				\quad \mathrm{C3_{y} = g^{(q_y(0))}}\\
				\quad \mathrm{C4_{y} = H(\mathtt{attr(y)})^{q_y(0)}}\\
			\end{rcases} {\mathrm{\forall\ y \in Y_{iv}}}\\
		\end{cases}$
		
		\item [] \textit{Phase 2}: In this phase, adversary $\mathcal{A}$ can continue to send secret key generation requests with the same restriction as in $Phase 1$, i.e., $\mathrm{a_j} \notin \mathcal{T}^*$.
		
		\item [] \textit{Guess Phase}: In this phase, adversary $\mathcal{A}$ outputs a guess $\mathrm{b\prime} \in \{0, 1\}$. If $\mathrm{b\prime} = \mathrm{b}$, the simulator $\beta$ will guess that $\mu = 0$ and $\mathrm{Z}_\mu = \mathrm{e(g, g)^{abc}}$, otherwise will guess that $\mu = 1$ and $\mathrm{Z}_\mu = \mathrm{e(g, g)}^{\theta}$. When $\mathrm{Z}_u = \mathrm{e(g, g)^{abc}}$ the simulator $\beta$ gives the perfect simulation and $\mathrm{c}_{\mathcal{T}^*}$ is a valid ciphertext. Therefore, the advantage of the adversary is 
		\begin{align*}
			\mathrm{Pr} [ \mathrm{b\prime} = \mathrm{b} \mid \mathrm{Z}_\mu = \mathrm{e(g, g)^{abc}} ] = \frac{1}{2} + \epsilon
		\end{align*}
		If $\mu = 1$ then $\mathrm{Z}_\mu = \mathrm{e(g, g)}^\theta$ and $\mathrm{c}_{\mathcal{T}^*}$ is random ciphertext for the adversary, and the adversary does not gain information about $\mathrm{m_b}$. Hence, we have
		\begin{align*}
			\mathrm{Pr} [ \mathrm{b\prime} \neq \mathrm{b} \mid \mathrm{Z}_\mu = \mathrm{e(g, g)}^{\theta} ] = \frac{1}{2}
		\end{align*}
		Since the simulator $\beta$ guesses $\mu\mathrm{\prime} = 0$ when $\mathrm{b\prime} = \mathrm{b}$ and $\mu\mathrm{\prime} = 1$ when $\mathrm{b\prime} \neq \mathrm{b}$, the overall advantage of $\beta$ to solve DBDH assumption is 
		\begin{align*}
		\dfrac{1}{2} \mathrm{Pr}[\mu\mathrm{\prime} = \mu \mid \mu = 0] + \dfrac{1}{2} \mathrm{Pr}[\mu\mathrm{\prime} = \mu \mid \mu = 1] - \dfrac{1}{2} = \dfrac{\epsilon}{2}
		\end{align*}
		If the adversary $\mathcal{A}$ has the above advantage $\epsilon$ to win the IND-sAtt-CPA game, the challenger can solve the DBDH assumption problem with $\epsilon/2$ advantage with the help of adversary $\mathcal{A}$. However, there are no effective polynomial algorithms which can solve the DBDH assumption problem with non-negligible advantage according to the DBDH assumption. Hence, the adversary cannot win the IND-sAtt-CPA game with the above advantage $\epsilon$, namely the adversary having no advantage to break through the proposed scheme.
	\end{itemize}
	
\section {\uppercase{Conclusion}}
\noindent
This paper presented a scheme to improve data privacy in DigiLocker by using CP-ABE. The scheme also proposed to prepone part of the encryption process to increase performance. This preponed process creates a token which can be reused later. The proposed scheme is proved to be secure against IND-sAtt-CPA game. The proposed scheme can further be enhanced by using homomorphic encryption which allows processing on encrypted data and using post-quantum ABE schemes, for both of which, though schemes exist but are still non-trivial and not practical. 

\bibliographystyle{apalike}
{\small
\bibliography{example}}

\begin{thebibliography}{}

\bibitem[Bethencourt et~al., 2007]{bethencourt2007ciphertext}
Bethencourt, J., Sahai, A., and Waters, B. (2007).
\newblock Ciphertext-policy attribute-based encryption.
\newblock In {\em 2007 IEEE symposium on security and privacy (SP'07)}, pages
  321--334. IEEE.

\bibitem[GoI, 2015]{digilocker}
GoI (2015).
\newblock Digilocker.
\newblock \url{https://digilocker.gov.in}.

\bibitem[Goyal et~al., 2006]{goyal2006attribute}
Goyal, V., Pandey, O., Sahai, A., and Waters, B. (2006).
\newblock Attribute-based encryption for fine-grained access control of
  encrypted data.
\newblock In {\em Proceedings of the 13th ACM conference on Computer and
  communications security}, pages 89--98.

\bibitem[Ibraimi et~al., 2009]{ibraimi2009efficient}
Ibraimi, L., Tang, Q., Hartel, P., and Jonker, W. (2009).
\newblock Efficient and provable secure ciphertext-policy attribute-based
  encryption schemes.
\newblock In {\em International Conference on Information Security Practice and
  Experience}, pages 1--12. Springer.

\bibitem[IETF, 2012]{oauth2}
IETF (2012).
\newblock The oauth 2.0 authorization framework.
\newblock \url{https://tools.ietf.org/rfc/rfc6749.txt}.

\bibitem[Ji et~al., 2014]{ji2014privacy}
Ji, Y.-m., Tan, J., Liu, H., Sun, Y.-p., Kang, J.-b., Kuang, Z., and Zhao, C.
  (2014).
\newblock A privacy protection method based on cp-abe and kp-abe for cloud
  computing.
\newblock {\em JSW}, 9(6):1367--1375.

\bibitem[MeitY, 2019]{dlts}
MeitY (2019).
\newblock Digital locker technical specification (dlts).
\newblock
  \url{https://img1.digitallocker.gov.in/assets/img/technicalspecifications-dlts-ver-2.3.pdf}.

\bibitem[UIDAI, 2009]{aadhaar}
UIDAI (2009).
\newblock What is aadhaar.
\newblock \url{https://uidai.gov.in/myaadhaar/about-your-aadhaar.html}.

\bibitem[Yacobi, 2002]{yacobi2002note}
Yacobi, Y. (2002).
\newblock A note on the bilinear diffie-hellman assumption.
\newblock {\em IACR Cryptology ePrint Archive}, 2002:113.

\bibitem[Zhang et~al., 2004]{zhang2004efficient}
Zhang, F., Safavi-Naini, R., and Susilo, W. (2004).
\newblock An efficient signature scheme from bilinear pairings and its
  applications.
\newblock In {\em International Workshop on Public Key Cryptography}, pages
  277--290. Springer.

\bibitem[Zhou and Huang, 2012]{zhou2012efficient}
Zhou, Z. and Huang, D. (2012).
\newblock Efficient and secure data storage operations for mobile cloud
  computing.
\newblock In {\em 2012 8th international conference on network and service
  management (cnsm) and 2012 workshop on systems virtualiztion management
  (svm)}, pages 37--45. IEEE.

\end{thebibliography}

\end{document}